\documentclass[aps,prl,preprint,groupedaddress,amsmath,amssymb,showpacs,showkeys]{revtex4}

\usepackage[english]{babel}
\usepackage[dviwin]{graphicx}
\usepackage[dvips]{color}
\usepackage{graphicx}
\usepackage{dcolumn}
\usepackage{bm}
\begin{document}


\title{Three-qubit dynamics of entanglement in the magnetic field}


\author{E. A. Ivanchenko}
\email{yevgeny@kipt.kharkov.ua}

\affiliation{Institute for Theoretical Physics, National Science Center
\textquotedblleft{}Institute of Physics and Technology\textquotedblright{},
   Kharkov, Ukraine}
\date{\today}

\begin{abstract}
A closed system of the equations for the local Bloch vectors and spin correlation
functions is obtained by decomplexification of the Liouville-von Neumann equation for
three magnetic qubits with the exchange interaction, that takes  place in an
arbitrary time-dependent external magnetic field.  The numerical comparative analysis
of entanglement is carried out depending on initial conditions and the magnetic field
modulation. The present study may be useful for analysis of interference experiments
and in the field of quantum computing.
\end{abstract}




\maketitle
\section{Introduction}
\indent The phenomenon of entanglement (inseparability) is one of the properties of
quantum systems. A quantum entanglement is the key source of such applications  as
superdense coding, teleportation,  quantum calculations researched by the quantum
information theory \cite {Galber, Dbou, Anie}. The most known and natural way
creating the entanglement   is the global unitary evolution caused by the interaction
between subsystems, which is controllable by external fields. The aim of this work
has been to make a comparative analysis of the measures of entanglement in  a
three-qubit system located in a magnetic field, because up to now contrary to
two-qubit system, there has been  no unified measure entanglement for three qubits.
\section{The Model Hamiltonian}
\indent The Hamiltonian of three coupled by exchange interaction magnetic qubits $e,
p, n $ (particles with spin 1/2)  placed in an external variable magnetic field $ \bm
{H} = (H_1, H_2, H_3) $ looks like
\begin{equation}
\hat{H}=h^e_is^e_i+h^p_is^p_i
+h^n_is^n_i+2J^{ep}s^e_is^p_i+2J^{en}s^e_is^n_i+2J^{pn}s^p_is^n_i,
\end{equation}
where $h^e_i, h^p_i, h^n_i $ are the Cartesian components of the external magnetic
field multiplied by gyromagnetic ratio of the  corresponding qubits;
 $s^e_i=\frac{1}{2}\sigma_i\otimes\sigma_0\otimes\sigma_0$,
 $s^p_i=\frac{1}{2}\sigma_0\otimes\sigma_i\otimes\sigma_0 $, $s^n_i=\frac{1}{2} \sigma_0\otimes\sigma_0\otimes\sigma_i $
  is the   matrix representation of spin operators of magnetic qubits;
    the Pauli matrices are equal to
  $ \sigma_0=\left(
\begin{array}{cc}
  1 & 0 \\
  0 & 1 \\
\end{array}
\right),  ~ \sigma_1=\left(
\begin{array}{cc}
  0 & 1 \\
  1 & 0 \\
\end{array}
\right), ~ \sigma_2=\left(
\begin{array}{cc}
  0 & -i \\
  i & 0 \\
\end{array}
\right), ~ \sigma_3=\left(
\begin{array}{cc}
  1 & 0 \\
  0 & -1 \\
\end{array}
\right)$; $\otimes$ is the symbol of direct product \cite{PL}; $J^{ep},J^{en},J^{pn}$
are the constants of isotropic exchange interaction between qubits;  the summation
over  $e, p, n $ is  absent.

\section{Decomplexification of the Liouville-von Neumann
equation}
 The Liouville-von Neumann
equation for the density matrix  $ \rho $, describing the dynamics of a three-qubit
system, looks like

\begin{equation}
  i\partial_t\rho=[\hat{H},\rho],~ \rho(t=0)=\rho_0.
\end{equation}
Let us present the solution of the equation (2) as
\begin{equation}
  \rho=\frac{1}{8}R_{\alpha\beta\gamma}\sigma_\alpha\otimes\sigma_\beta\otimes\sigma_\gamma,~
  \rho^+=\rho,~R_{000}=1,~Sp\rho=1.
\end{equation}
 Hereinafter summation is taken over  the repeating  Greek indices from zero up to four,
and over Latin indices from one up to three. The three coherence vectors  (the Bloch
vectors)  widely used in  the magnetic resonance theory, are written as
\begin{subequations}
\begin{equation}
R_{i00}=Sp\rho\sigma_i\otimes\sigma_0\otimes\sigma_0,
\end{equation}
\begin{equation}
R_{0i0}=Sp\rho\sigma_0\otimes\sigma_i\otimes\sigma_0,
\end{equation}
\begin{equation}
R_{00i}=Sp\rho\sigma_0\otimes\sigma_0\otimes\sigma_i.
\end{equation}
\end{subequations}
They  characterize the local properties of individual qubits, whereas the following
tensors
\begin{subequations}
\begin{equation}
R_{kq0}=Sp\rho\sigma_k\otimes\sigma_q\otimes\sigma_0,
\end{equation}
\begin{equation}
R_{k0q}=Sp\rho\sigma_k\otimes\sigma_0\otimes\sigma_q,
\end{equation}
\begin{equation}
R_{0kq}=Sp\rho\sigma_0\otimes\sigma_k\otimes\sigma_q,
\end{equation}
\begin{equation}
R_{kql}=Sp\rho\sigma_k\otimes\sigma_q\otimes\sigma_l
\end{equation}
\end{subequations}
describe the spin correlations.  \\
\indent The length of the generalized Bloch vector $ b $   is conserved under unitary
evolution.
\begin{equation}
b =\sqrt{R^2_{\alpha\beta\gamma}}.
\end{equation}
The Liouville-von Neumann equation accepts the real form in terms of the functions $R
_ {\alpha\beta\gamma} $ as  closed system of 63 differential equations  for the local
Bloch vectors  and spin correlation functions
\begin{subequations}
\begin{equation}
\partial_t R_{q00} =
 \varepsilon_{ilq}h^e_iR_{l00}+\varepsilon_{mlq}(J^{ep}R_{lm0}+J^{en}R_{l0m}),
\end{equation}
\begin{equation}
  \partial_t R_{0q0} =
  \varepsilon_{ilq}h^p_iR_{0l0}+\varepsilon_{mlq}(J^{ep}R_{ml0}+J^{pn}R_{0lm}),
\end{equation}
\begin{equation}
  \partial_t R_{00q} =
  \varepsilon_{ilq}h^n_iR_{00l}+\varepsilon_{lmq}(J^{en}R_{lom}+J^{pn}R_{0lm}),
\end{equation}
\begin{eqnarray}
\partial_t R_{qk0}&=
 \varepsilon_{ilq}h^e_iR_{lk0}+\varepsilon_{imk}h^p_iR_{qm0}+J^{ep}\varepsilon_{kmq}(R_{m00}-R_{0m0})\nonumber\\
 &\quad +J^{en}\varepsilon_{lmq}R_{mkl}+J^{pn}\varepsilon_{lmk}R_{qml},
  \end{eqnarray}
\begin{eqnarray}
 \partial_t R_{q0k} &=
 \varepsilon_{ilq}h^e_iR_{l0k}+\varepsilon_{imk}h^n_iR_{q0m}+J^{en}\varepsilon_{qmk}(R_{00m}-R_{m00})\nonumber\\
 &\quad +J^{ep}\varepsilon_{lmq}R_{mlk}+J^{pn}\varepsilon_{lmk}R_{qlm},
 \end{eqnarray}
\begin{eqnarray}
 \partial_t R_{0qk} &=
 \varepsilon_{ilq}h^p_iR_{0lk}+\varepsilon_{imk}h^n_iR_{0qm}+J^{pn}\varepsilon_{qmk}(R_{00m}-R_{0m0})\nonumber\\
 &\quad +J^{ep}\varepsilon_{lmq}R_{lmk}+J^{en}\varepsilon_{lmk}R_{lqm},
 \end{eqnarray}
 \begin{eqnarray}
 \partial_t R_{qkl} &=
 \varepsilon_{imq}h^e_iR_{mkl}+\varepsilon_{imk}h^p_iR_{qml}+\varepsilon_{iml}h^n_iR_{qkm}+J^{ep}\varepsilon_{kmq}(R_{m0l}-R_{0ml})\nonumber\\
 &\quad  +J^{en}\varepsilon_{qml}(R_{0km}-R_{mk0})+J^{pn}\varepsilon_{kml}(R_{q0m}-R_{qm0})
 \end{eqnarray}
 \end{subequations}
 for the set of initial  conditions.\\
   \indent  In system (7), assuming, for example, $J ^ {en} =0 $,
$J ^ {pn} =0 $, we get   the closed system of  equations for the description of
two-qubit dynamics
\begin{subequations}
\begin{equation}
\partial_t R_{q0} =
 \varepsilon_{ilq}h^e_iR_{l0}+\varepsilon_{mlq}J^{ep}R_{lm},
\end{equation}
\begin{equation}
  \partial_t R_{0q} =
  \varepsilon_{ilq}h^p_iR_{0l}+\varepsilon_{mlq}J^{ep}R_{ml},
\end{equation}
\begin{equation}
\partial_t R_{qk}=
 \varepsilon_{ilq}h^e_iR_{lk}+\varepsilon_{imk}h^p_iR_{qm}+J^{ep}\varepsilon_{kmq}(R_{m0}-R_{0m}),
 \end{equation}
where $ R_{q0}=Sp\rho\sigma_q\otimes\sigma_0$,  $
R_{0q}=Sp\rho\sigma_0\otimes\sigma_q$, $R_{kq}=Sp\rho\sigma_k\otimes\sigma_q$.
 \end{subequations}\\
The concrete calculations will be carried out for the following initial conditions.
\\
 The fully separable state  (S) is \\
 \begin{equation}
\mid S>=\mid 111>,
 \end{equation}
 the biseparable state (BS) is \cite{DCT}
\begin{equation}
\mid BS>=\frac{1}{\sqrt{2}}(\mid 001>+\mid 010>),
\end{equation}
the Greenberger-Horne-Zeilinger maximally  entangled state  (GHZ) is
\begin{equation}
\mid GHZ>=\frac{1}{\sqrt{2}}(\mid 000>+\mid 111>),
\end{equation}
the Werner entangled state  (W) is
\begin{equation}
\mid W>=\frac{1}{\sqrt{3}}(\mid 001>+\mid 010>+\mid 100>),
\end{equation}
the mixed state (Mix) \cite{Wei} is
\begin{equation}
\rho_0=x\mid GHZ><GHZ\mid\ +\frac{1-x}{2}(\mid W><W\mid + \mid V><V\mid),
 \end{equation}
 where
\begin{equation}
\mid V>=\frac{1}{\sqrt{3}}(\mid 110>+\mid 101>+\mid 011>), 1/3<x\leq 1.
\end{equation}
The length of the generalized  Bloch vector  (6) for pure states  (9 -12) is equal to
 $ \sqrt{7}$, and for the mixed state (13) at $x= \frac{2}{3}$ it is equal to  $ \sqrt{3}$.

\section{Measures of global entanglement in the three-qubit system}

     We shall define the measures of entanglement
   of three qubits \cite {SM} on solutions of the system (7), by introducing the tensors
   of two-partcle entanglement:

\begin{subequations}
 \begin{equation}
m_{ij0}=R_{ij0}-R_{i00}R_{0j0},
\end{equation}
\begin{equation}
m_{i0j}=R_{i0j}-R_{i00}R_{00j},
\end{equation}
\begin{equation}
m_{0ij}=R_{0ij}-R_{0i0}R_{00j}.
\end{equation}
\end{subequations}
The tensors (15) are equal to zero, if the two-particle correlation functions (5a,
5b, 5c) are factorized in terms of the Bloch vectors  (4). With the help of these
tensors we shall define the measure of two-particle entanglement in two-qubit system
as

\begin{equation}
m=m_{ij}^2,
\end{equation}
where  $m_{ij}=R_{ij}-R_{i0}R_{0j}$.
 The three-particle entanglement tensor $m _ {ijk} $
is obtained by subtraction from three-particle spin correlation function $R _ {ijk} $
(5d)of all products of the Bloch vectors and   the tensors of lower order

\begin{equation}
m_{ijk}=R_{ijk}-R_{i00}m_{0jk}-R_{0j0}m_{i0k}-R_{00k}m_{ij0}-R_{i00}R_{0j0}R_{00k}.
\end{equation}
 This decomposition can be considered as a kind of cluster expansion in statistical
mechanics \cite {F}. In terms of these tensors the measure of three-particle
entanglement becomes \cite {SM}

\begin{equation}
m_{SM}=m_{ijk}^2.
\end{equation}

This measure is equal to zero when the tensor $R _ {ijk} $ can be expressed through
two-particle entanglements (15) and the local Bloch vectors. It  is also
applicable  for  pure as well as for mixed states.\\
 The  concurrence \cite {CMB} depends on all six reduced density matrices, and for
pure states it  can be represented  as

\begin{equation}
C_3=\frac{1}{\sqrt{2}}\sqrt{6-[\frac{9}{4}+R_{i00}^2+R_{0i0}^2+R_{00i}^2+
\frac{1}{4}(R_{mn0}^2+R_{m0n}^2+R_{0mn}^2)]}.
\end{equation}\\
The measures $m _ {SM} $ and $C_3 $ are not normalized to 1. \\
 The measure of global entanglement \cite {MW}
in the form \cite {Br}  is expressed through the reduced matrices of individual
qubits $\rho_e=Sp_{pn}\rho$, $\rho_p=Sp_{en}\rho$, $\rho_n=Sp_{ep}\rho$  according to
the formula
\begin{equation}
m_B=1-\frac{R_{i00}^2+R_{0i0}^2+R_{00i}^2}{3}.
\end{equation}
  For the initial GHZ
state of three qubits the three-tangle measure, expressed through the Cayley
hyperdeterminant \cite {CKW},  is equal to

\begin{equation}
m_K=4\rho_{11}\rho_{88},
\end{equation}
where
  the diagonal density matrix elements determine
 population probabilities\\
\begin{equation}
\rho_{11}=\frac{1}{8}(R_{300}+R_{030}+R_{003}+R_{330}+R_{303}+R_{033}+R_{333}+1),
\end{equation}
\begin{equation}
\rho_{88}=\frac{1}{8}(- R_{300}-R_{030}-R_{003}+R_{330}+R_{303}+R_{033}-R_{333}+1).
\end{equation}
 The measure of global entanglement is not equal to zero for fully
 inseparable pure states \cite {LS} and  is determined by the expression
\begin{equation}
m_L=\sqrt[3]{(1-R_{i00}^2)(1-R_{0i0}^2)(1-R_{00i}^2)}.
\end{equation}
  This measure is equal to one for initial GHZ state and to 8/9 for the W state,
  and it takes the same
values for the $m_B $ measure.
\section{Numerical results}
Let us consider two cases, when the external magnetic fields dependent on the
dimensionless time
 $\tau=\omega t$:\\
 the resonant ( relative to $e $ qubit) circularly  polarized field (R) is
\begin{equation}
\bm{H}=-(\frac{\omega_1}{\omega}cos\tau,~-\frac{\omega_1}{\omega}sin\tau,~\frac{\omega_0}{\omega}),
\end{equation}
 and  the nonresonant  circularly  polarized field (NR) is
\begin{equation}
\bm{H}=-(\frac{\omega_1}{\omega}cos\tau,~\frac{\omega_1}{\omega}sin\tau,~\frac{\omega_0}{\omega}),
\end{equation}
where $\omega$ is the frequency of the external field , $\frac{\omega_1}{\omega}$ and
$\frac{\omega_0}{\omega}$ are the dimensionless amplitudes of  transverse and
longitudinal field correspondingly. Let us assume, that the fields, operating on the
qubits $e $, $p $ and $n $, are equal  to $ \bm {h^e} = \bm {H} $, $ \bm {h^p} =2\bm
{H} $ and $ \bm {h^n} =4\bm {H} $, respectively.  To perform the numerical
simulation, we have chosen
 the following parameters in units  of $2 \pi \times 100 $ MHz, $\omega=\omega_0=1$.
  It corresponds to the longitudinal field
 $H_3=2.3487$ Ts for the proton resonance of $e$ qubit.
 The  exchange constants are equal to $J ^ {ep} =-0.2
$; $J ^ {en} =-0.1 $; $J ^ {pn} =-0.3$ and $\omega_1=0.3$  in the same units.
  In figures 1 and 2 the solid line corresponds to the NR field, and dashed line - to the  R field.
  \begin{figure}
 \includegraphics[width=7in]{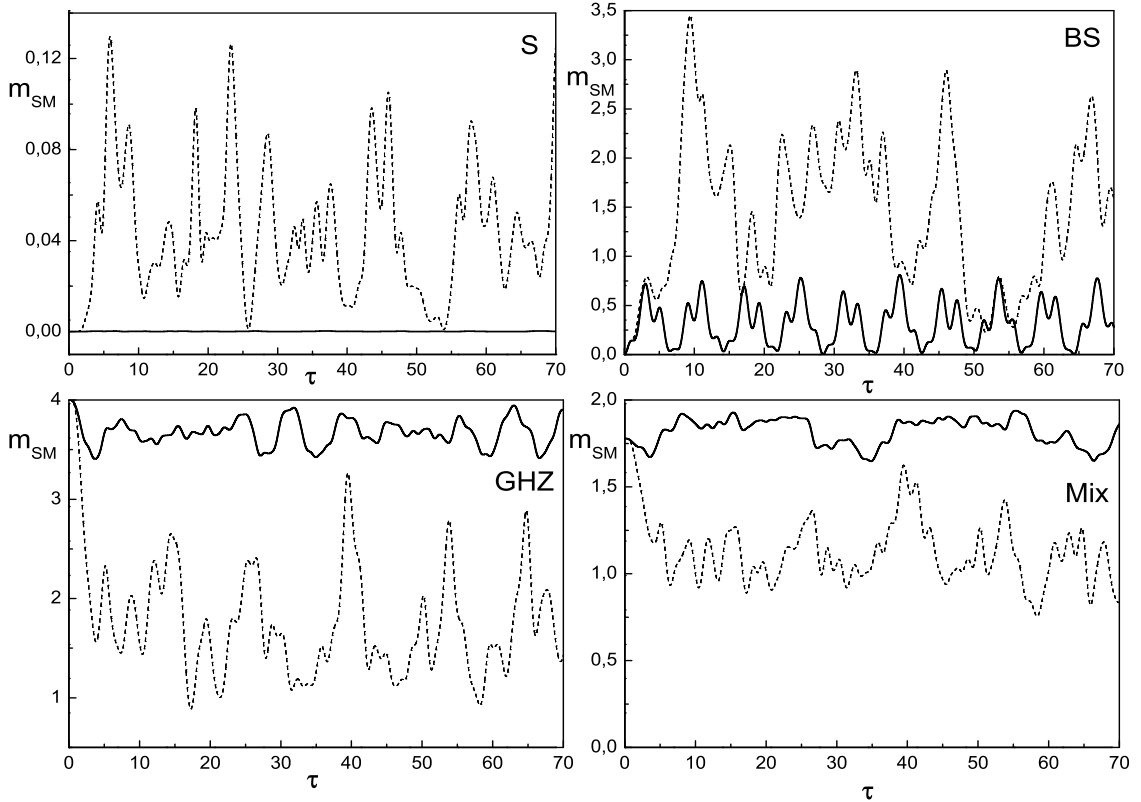}
 \caption{\label{Mahler}Entanglement dynamics    $m _ {SM}$ versus initial conditions}
 \end{figure}\\
  \indent {\it{The S state}}.
The measure $m _ {SM} $   is not  sensitive to the NR field (amplitude of
fluctuations is less than 0.001), but it shows a resonant behavior in the R field
(Fig.1).
  The measures  $C_3, m_B,
m_L $ display irregular fluctuations with a low amplitude in the  NR  field and with
a high amplitude in the R  field.  Fig. 2 shows the dynamics of  measure $ m_L $.
 It can be seen that the behavior of  measures $m_{SM}$ and $ m_L $ is similarly.
\begin {figure}
 \includegraphics[width=7in]{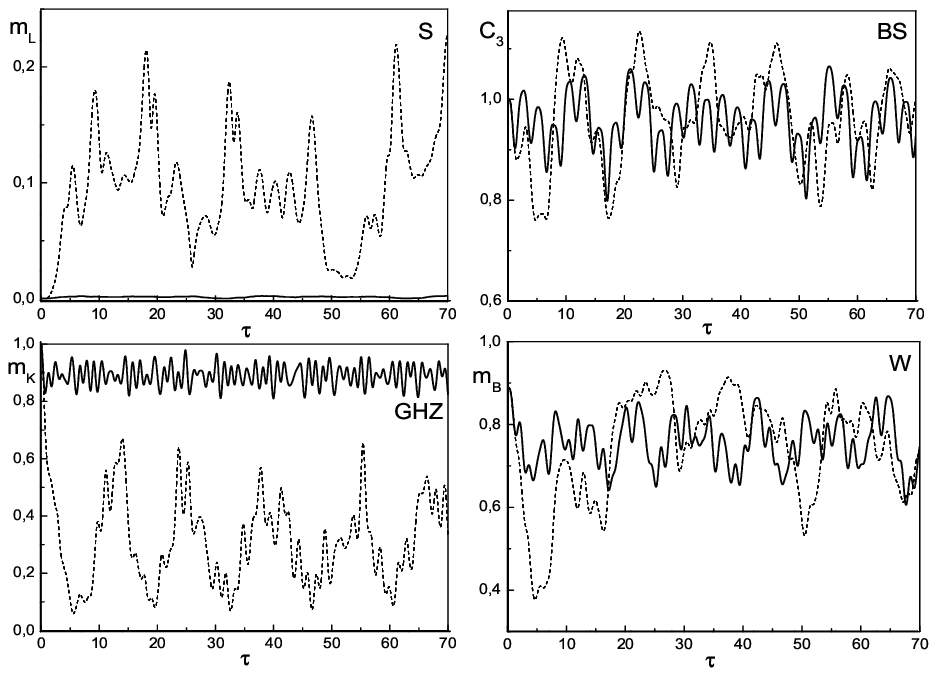}
 \caption {\label {L,C,K,B} Entanglement dynamics
  of the measures $m_L $, $C_3 $, $m_K $,
 $m_B $ versus initial conditions}
 \end {figure} \\
  \indent {\it {The BS state}}. The measure $m _ {SM} $ has a  periodic character in the NR  field
  and shows
irregular fluctuations with a high amplitude in the R  field. The amplitude of
fluctuations in the R  field is 3 times larger than in the NR  field.
  The entanglement measures  $C_3, m_B, m_L $ display irregular
fluctuations with a high amplitude, and  they are also little different for the R and
NR fields.
 The measure $ C_3 $ is shown in Fig. 2. \\
\indent {\it {The GHZ state}}. The measure $m _ {SM} $  is well distinct  for R and
NR fields.
 The measures  $C_3, m_B, m_K, m_L $ exhibit  low-amplitude
fluctuations in the NR  field. In the R  field the entanglement is less than in the
NR field at all times, but the occurring fluctuations have a high amplitude.  Fig. 2
shows the dynamics of  measure
 $ m_K $. \\
\indent {\it {The W state}}. The measure $m _ {SM} $ has an oscillatory character at
times of $0 <\tau <15 $  and  also it is also practically indiscernible for R and NR
fields. For $ \tau> 20 $ the amplitude of fluctuations in the R  field is 3 times
higher than in the NR  field (it is not shown in Fig. 1).
 The entanglement measures  $C_3, m_B, m_L $ display
a qualitatively  similar behavior in  both  the R and NR  field.
  Fig. 2 shows the dynamics of  measure $ m_B $. \\
 \indent {\it {The Mix  state}}.
  It can be  seen from  Fig. 1, that the measure $m _ {SM} $
 insignificantly changes in comparison with the initial value in the NR  field, while the R  field
 appreciably reduces the entanglement. \\
 \indent
  As a matter of fact, the solution of  equation (2) for a constant external field
  looks like
  \begin{equation}
  \rho=exp(-i \hat{H} t)\rho_0 exp(i \hat{H} t).
\end{equation}
Evidently,  if  $[\rho_0,\hat{H}]=0$,  then  $\rho=\rho_0$ for all $t$. Thus the
value of an arbitrary function dependent on solutions of  equation (2) will be equal
to the initial value. In  other words, the density matrix will be fixed for this
Hamiltonian.
  The influence of time-dependent magnetic field is
insignificant  for $ \omega_0 >> \omega_1 $, therefore all entanglement  measures
 are
constant and equal to the  initial entanglement. \\
\begin{figure}
 \includegraphics [width=3.4in] {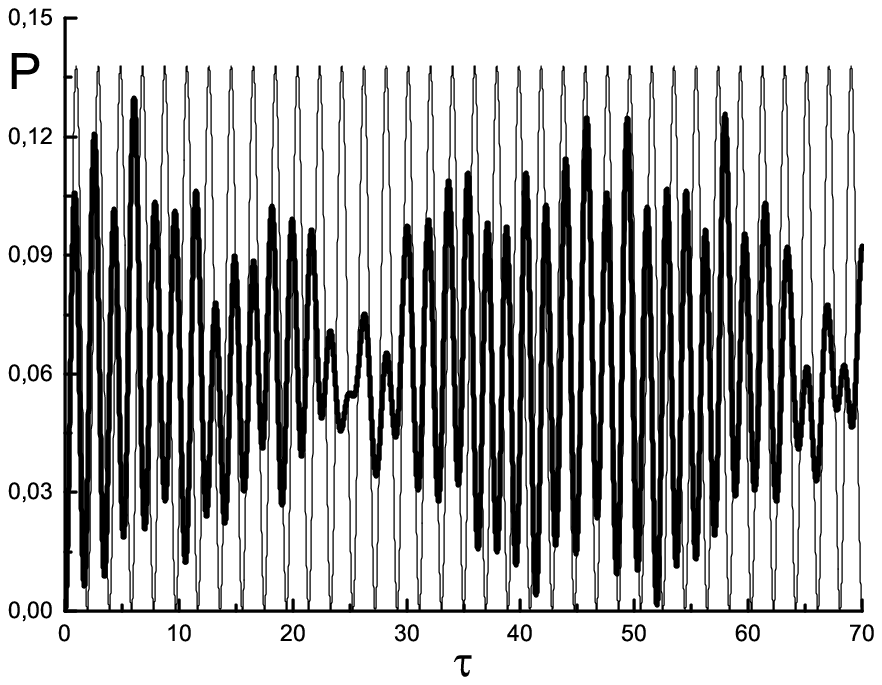}
 \caption {\label {Prob} The influence of $e $, $p $ qubits  as fluctuators (a thick line)
  on  spin flip probability of the free $n $ qubit $n $ (a thin line) in the R  field}
 \end{figure}
 In the  model considered, a decrease  in the transverse  field  amplitude leads to
  the decrease  in the amplitudes of fluctuations for all entanglement measures,  regardless
  of the  initial conditions. \\
 \indent We  shall consider the influence of  qubits $e, p $ as two fluctuators \cite {Alt}
 on the $n $ qubit dynamics.
  From system (7c) it follows
   for $J ^ {en} =0 $, $J ^ {pn} =0 $, that the spin flip probability
  of the
  free qubit $n $ from an initial state
  $R _ {001} (0) =0 $, $R _ {002} (0) =0 $, $R _ {003} (0) =1 $  is equal to \cite {EAI}
\begin {equation}
  P = \frac {1-R _ {003}} {2}.
\end {equation}
   Fig. 3 shows the spin flip probability of $n $ qubit
    in the resonant field
  and the dependence of  this probability  on the field deformation   caused
 by the presence of qubits $e, p $ in
 the initial S state. It can be seen  that the Rabi oscillations transform to beats.
  This is in qualitative  accordance with the results of work \cite {Alt}.
\section {Conclusion}
The closed system of
  equations for the local Bloch vectors  and spin correlation functions
  for three magnetic qubits with the exchange interaction, that takes place in any
  time-dependent  external magnetic field has been  derived.
  The numerical comparative analysis of entanglement measures has been made,  depending on initial
   conditions and  the magnetic   field modulation.\\
   \indent From the experimental point of view, the measure $m_K $ is
   more acceptable
as it is expressed through the populations $ \rho _ {11} $, $ \rho _ {88} $.
 The  measures $C_3$  and  $ m_{SM} $ for GHZ conditions which also show sharp difference
for R and NR fields but they   depend in a complicated manner on all elements
of the  density matrix $\rho$ . \\
The present study will permit one to bring closer the theoretical
   results \cite {SM, CMB, MW, Br, CKW, LS}
    to the possibility  of experimental
   corroboration.\\
   \indent The proposed approach can be realized without any specific difficulties
    for the
   four-qubit system the detailed dynamics of which is described by 255 equations with
   the length of the generalized Bloch vector for pure states $ b=\sqrt{15} $.

\begin {acknowledgments}
The author is grateful to Zippa Anna Anikeyevna for fruitful
discussion
 and constant invaluable support. 
\end {acknowledgments}



\thebibliography{99}
\bibitem{Galber} G. Alber, T. Beth, M. Gorodecki, R. Gorodecki, M. Rotteler, H.
Weinfurter, R. Werner, and Zeilinger \textit{Quantum Information: An Introduction to
Basic Theoretical Concepts and Experiments} (Springer Verlag, 2001).
\bibitem{Dbou} D. Bouwmeester, A. Ekert, and Zeilinger \textit{The Physics
of Quantum Information} (Springer Verlag, 2000).
\bibitem{Anie} A. Nielsen and I. Chuang, \textit{Quantum
Computation and Quantum Information} (Cambridge University Press, 2000).
\bibitem{PL} P. Lankaster \textit{Theory of matrices} (Academic Press, New York-London, 1969).
\bibitem{DCT}  W. D\"{u}r, J. I. Cirac, and R. Tarrach.  Phys. Rev. Lett. {\bf 83}, 3562 (1999).
\bibitem{Wei} Tzu-Chieh Wei and Paul M. Goldbart, Phys. Rev. A {\bf 68}, 042307 (2003).
 \bibitem{SM} J. Schlienz and G.Mahler, Phys. Rev. A {\bf 52}, 4396-4404 (1995).
\bibitem{F} S. Fujita. \textit{Introducion to Non-Equilibrium Quantum Statistical Mechanics} (Krieger,
 FL, 1983).
\bibitem{CMB}  A. R. R. Carvalho, F. Mintert, and A. Buchleitner,  Phys. Rev. Lett. {\bf 93},
230501 (2004).
 \bibitem{MW} D. A. Meyer and N. R. Wallach, J. Math. Phys. {\bf 43}, 4273 (2002).
 \bibitem{Br} G. K. Brennen. Quant. Inf. and Comput.,  {\bf 3}, 616 (2003).
 \bibitem{CKW} V. Coffman, J. Kundu , and W.K. Wootters,  Phys. Rev. A {\bf 61}, 052306 (2000).
\bibitem{LS} P. J. Love \textit{et al.,} quant-ph/0602143 (2006).
\bibitem{Alt} Y.M. Galperin \textit{et al.,} Europhys. Lett. {\bf 71}, 21 (2005);
 cond-mat/0501455 (2005).
 \bibitem{EAI} E. A. Ivanchenko,  Fiz. Nizk. Temp. \textbf{31},  761 (2005).
 \end{document}